
\input phyzzx

\nonstopmode
\twelvepoint
\nopubblock
\overfullrule=0pt
\tolerance=5000

\line{\hfill }
\line{\hfill IASSNS-HEP-93/23}
\line{\hfill hep-ph/9304318}
\line{\hfill March 1993}

\titlepage
\title{Beyond the Standard Model\foot{Invited talk at
PASCOS Conference, Nov. 1992, Berkeley }}

\author{Frank Wilczek\foot{Research supported in part by DOE grant
DE-FG02-90ER40542.~~~WILCZEK@IASSNS.BITNET}}
\vskip.2cm
\centerline{{\it School of Natural Sciences}}
\centerline{{\it Institute for Advanced Study}}
\centerline{{\it Olden Lane}}
\centerline{{\it Princeton, N.J. 08540}}

\abstract{The standard model of particle physics is marvelously
successful.  However, it is obviously not a complete or final theory.
I shall argue here that the structure of the standard model
gives some quite concrete, compelling hints regarding what lies beyond.}

\REF\drw{The reader may wish to compare S. Dimopoulos, S. Raby, F. Wilczek
{\it Physics Today} Oct. 1991, p. 25ff for a related discussion and
additional  references.}

\REF\aach{The best source for up-to-date information
and references is undoubtedly review talks at the standard international
conferences.  The recent (June 1992) Aachen conference {\it Twenty Years
of QCD\/}, ed. Zwirner, should be especially useful when it appears.}

\REF\lgt{Recently there has been rather spectacular progress in lattice
QCD, so anything other than the latest preprints does not represent the
state of the art.  See especially
G. Lepage, P. Mackenzie, {\it On the Viability of Lattice Gauge
Theory\/} (Cornell preprint, 1992) and F. Butler, H. Chen, J. Sexton,
A. Vaccarino, D. Weingarten {\it Hadron Mass Predictions of the
Valence Approximation to Lattice QCD\/} IBM report RC 19817 (81402).}

\REF\gegl{H. Georgi, S. Glashow {\it Phys. Rev. Lett}. {\bf 32}, 438
(1974).}

\REF\geo{H. Georgi, in {\it AIP Conference Proceedings\/}
{\bf 23}, ed. C. Carlson (AIP Press, 1975).}

\REF\gre{An accessible guide to string model building is
B. Greene, {\it Lectures on String Theory in Four Dimensions},
Cornell preprint CLNS 91-1046 (1991), to  appear in the Proceedings
of the 1990 Trieste Summer School on High Energy Physics and
Cosmology.}

\REF\afgt{D. Gross, F. Wilczek {\it Phys. Rev. Lett}. {\bf 30}, 1343
(1973);  {\it Phys. Rev}. {\bf D8}, 3633 (1973); H. D. Politzer
{\it Phys. Rev. Lett}. {\bf 30}, 1346 (1973).}

\REF\para{R. Hughes {\it Phys. Lett}. {\bf B97}, 246 (1980); {\it Nucl.
Phys}. {\bf B186}, 376 (1981); N. K. Nielsen {\it Am. J. Physics\/}
{\bf 49}, 1171 (1981).}

\REF\gqw{H. Georgi, H. Quinn, S. Weinberg {\it Phys. Rev. Lett}.
{\bf 33}, 451 (1974).}

\REF\pdec{R. Becker-Szendy {\it et al.}, {\it Phys. Rev}. {\bf D42},
2974 (1990); Particle Data Group, {\it Phys. Lett}. {\bf  B239}, 1 (1990).}

\REF\ferr{A very useful reprint collection with some historical
commentary is S. Ferrara, ed. {\it Supersymmetry\/} (2 vols.)
(North-Holland/World Scientific, Singapore 1987).}

\REF\srun{S. Dimopoulos, S. Raby, F. Wilczek {\it Phys. Rev}.
{\bf D24}, 1681 (1981); W. Marciano and G. Senjanovic
{\it Phys. Rev}. {\bf D25}, 3092 (1982); M. Einhorn and D. Jones
{\it Nucl. Phys}. {\bf B196}, 475 (1982).}

\REF\scalc{U. Amaldi, W. de Boer, H. Furstenau {\it Phys. Lett}.
{\bf B260}, 447 (1991); P. Langacker, M. Luo
{\it Phys. Rev}. {\bf D44}, 817 (1991).  See also
A. Sirlin {\it Phys. Lett}. {\bf B232}, 123 (1989);
J. Ellis, S. Kelley,
D. Nanopoulos {\it Phys. Lett}. {\bf B249}, 441 (1990).}

\REF\ccexpt{For a proper discussion of the data, see the
references quoted in [\scalc ].}

As befits a wise and efficient organizer,
Bernard Sadoulet called
to ask me to give this talk (with the assigned title)
long in advance of the conference.  Thus when I accepted
his invitation
it was
with a certain sense of unreality.
Only when the time came
to prepare, did I realize what a difficult chore it was that
I had taken on.  For, first of all, there have
been many many talks at previous conferences on the same subject
(that is, with the same assigned title);
and secondly, it is not a subject.
The first factor
makes it a challenge to say anything fresh; but
fortunately the second permits considerable flexibility.

What I decided to do, realizing that I would face a mixed audience
including many astronomers and specialists in general relativity, was
to try to convey in a simple but honest way the most compelling ideas
I know that lead one to concrete expectations for physics beyond the
standard model.  And in judging what was compelling, I tried to put
myself into the position of an intelligent and sympathetic but
properly skeptical physicist from outside particle physics.  If I were
such a person, who did not frequently hear particle physics talks, I
would not necessarily want to hear what the speaker considered the
{\it newest\/} or {\it most exciting\/} ideas -- that is, in
practice, whatever
the speaker has been working on for the last few months --
but rather the {\it best\/} and {\it most compelling\/}
ideas, which might be new and exciting
to me, and in any case would be
far less likely to prove transient or false.

Accordingly what follows contains neither new analysis of
experimental data nor ambitious new theoretical ideas.
It is mainly a record of
my judgement of what the central clues for
physics beyond the standard model are,
and an attempt at some pedagogy.
Experts looking for the latest bounds on Higgs, top, or
neutrino masses should look elsewhere.
So too should those interested in the latest wrinkles in superstring
or technicolor model building.  However even experts
might profit by stepping back occasionally for perspective,
and I'll try to keep it interesting for them by throwing in a few
provocations [\drw ].

\bigskip

{\bf Critique of the Standard Model}

The standard model of particle physics is based upon the
gauge groups SU(3)$\times$SU(2)$\times$U(1) of strong, electromagnetic
and weak interactions acting on the quark and lepton multiplets
as shown in Figure 1.

In this Figure I have depicted only
one family (u,d,e,$\nu_e$) of quarks and leptons;
in reality there seem to be three families which are mere copies of
one another as far as their interactions with the gauge bosons are
concerned, but differ in mass.
Actually in the Figure
I have ignored masses altogether, and allowed myself the convenient
fiction of pretending that the quarks and leptons have a definite
chirality -- right- or left-handed -- as they would if they were
massless.   (The more precise statement, valid when masses
are included, is that the gauge bosons
couple to currents of definite chirality.)  The chirality is indicated
by a subscript R or L.  Finally the little number
beside each multiplet is its assignment under the
U(1) of hypercharge, which is the average of the electric charge of the
multiplet.  (The physical photon is a linear
combination of the diagonal generator of SU(2) and the hypercharge
gauge bosons.  The physical Z boson is the orthogonal combination.)

\parskip=0pt{
$$\matrix{\rm SU(3)& \times& \rm SU(2)& \times& \rm U(1)\cr
\rm 8\ gluons&&\rm W^\pm,Z&&\gamma\cr}$$
\vglue-10pt
\hbox{$\qquad\qquad\qquad\qquad
\qquad\qquad\qquad\qquad\qquad
\ \ \ \ \ \uparrow\kern-6pt\underline{\ \ \ \ \ \
\ \ \ \ \ }\kern-6pt\uparrow$}

\moveright 3.65in \vbox{\offinterlineskip
\hbox{mixed}}

$\ \ \qquad \qquad
 \qquad \ \ \ \ \ \ \ \ \ \ {\rm SU(3)}$
\vglue-5pt
$\ \ \qquad \qquad
 \qquad \ \ \ \ \ \ \ \ \ \ \longleftrightarrow$
\vglue-18pt
$${\rm SU(2)}\updownarrow \pmatrix{\rm u{^r_L}&\rm u{^w_L}&\rm
u{^b_L}\cr
\rm d{^r_L}&\rm d{^w_L}&\rm d{^b_L}\cr}  {1\over 6} \qquad
\matrix{\rm (u{^r_R}&\rm u{^w_R}&\rm u{^b_R}){2 \over 3}\cr
\rm (d{^r_R}&\rm d{^w_R}&\rm d{^b_R})-{1 \over 3}\cr}$$

$$\left({\nu_{\rm L} \atop e_L} \right) -{1 \over 2} \ \ \ \ {\rm e_R}
-1$$

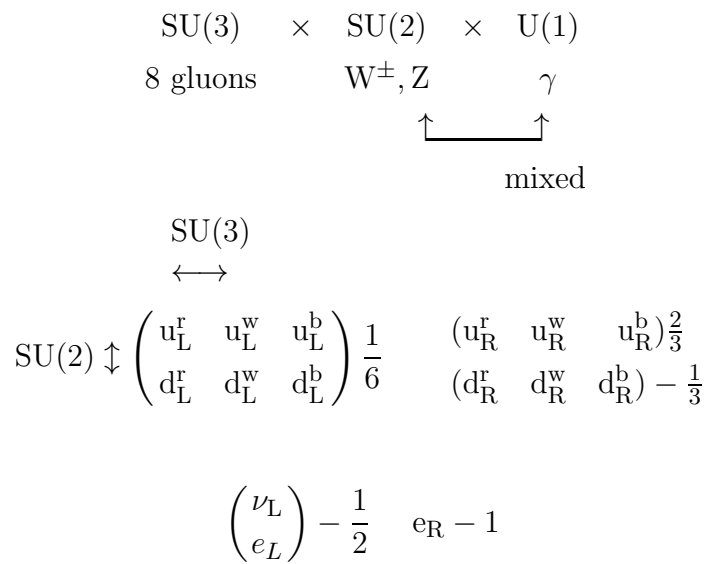
\centerline{\bf FIGURE 1}

\noindent{\it Figure 1 - The gauge groups of the standard model, and the
fermion
multiplets with their hypercharges.}
\bigskip}

Figure 1, properly understood -- that is, the standard model --
describes
a tremendous chunk of physics.  The strong interactions responsible
for the structure of nucleons and nuclei, and for most of what happens
in high energy collisions; the weak interactions responsible for
nuclear transmutations; and the electromagnetic interactions responsible
in Dirac's phrase for ``all of chemistry and most of physics'' are all
there, described by mathematically precise and indeed rather similar
theories of vector gauge particles interaction with spin-$1\over 2$
fermions.  The standard model
provides a remarkably compact description of all this.  It is
also a remarkably successful description,
with its fundamentals having now been
vigorously and rigorously tested in many experiments, especially at
LEP.  Precise quantitative comparisons between theory and
experiment are nothing new for QED and the weak interactions, but
if you haven't been paying attention you may not be aware that the
situation for QCD has improved dramatically in the last
few years [\aach ].
For example phenomenologists
now debate over the third decimal place in
the strong coupling constant, experiments
are now routinely sensitive to two-loop and
even three-loop QCD effects, and recent lattice gauge simulations are
achieving 10\% or better accuracy in the spectrum
both for heavy quark and
for light quark systems [\lgt ].

While
little doubt can remain
that the standard model is essentially
correct, a glance at Figure 1 is enough to reveal that it is
not a complete or final theory.
The fermions fall into apart into five lopsided pieces with
peculiar hypercharge assignments; this pattern needs to be
explained.  Also the separate gauge theories, which as I mentioned
are mathematically similar, are fairly begging to be unified.
Let me elaborate a bit on this.
The
SU(3) of strong interactions is, roughly speaking,
an extension of QED to three new types of charges, which in the
QCD context are called colors (say red, white, and blue).
QCD contains eight different gauge boson, or color gluons.
There are six possible gauge bosons
which transform one unit of any color charge into
one unit of any other, and two photon-like gauge bosons
that sense the colors.  An important subtlety which emerges simply from
the mathematics and which will play an
important role in our further considerations is that
there are two rather than three color-sensing gauge bosons.  This
is because the
linear combination which couples to all three color charges equally
is not part of SU(3).  Similarly the SU(2) of weak interactions is the
theory of two colors (say green and purple) and features three
gauge bosons: the weak color changing ones, which
we call W$^+$, W$^-$, and the weak color-sensing
one that mixes with the
U(1) hypercharge
boson to yield Z and the photon $\gamma$.

\bigskip
{\bf Unification: quantum numbers}
\medskip

Given that the strong interactions are governed by transformations
among three colors, and the weak by transformations between
two others, what
could be more natural than to embed both theories
into a larger theory of transformations among all five colors?
This idea has the additional attraction that an extra
U(1) symmetry commuting with the strong SU(3) and weak
SU(2) symmetries automatically appears,
which we can attempt to identify with the remaining gauge symmetry of
the standard model, that is
hypercharge.  For while in the separate SU(3) and SU(2) theories we
must throw out the two gauge bosons which couple respectively to
the color combinations R+W+B  and G+P, in the SU(5) theory we only
project out R+W+B+G+P, while the orthogonal
combination (R+W+B)-${3\over 2}$(G+P) remains.

Georgi and Glashow [\gegl ]
originated this line of thought, and showed how
it could be used to bring some order to the quark and lepton sector,
and in particular to
supply a satisfying explanation of the weird hypercharge assignments
in the standard model.  As shown in Figure 2, the five scattered
SU(3)$\times$SU(2)$\times$U(1) multiplets get organized into just two
representations of SU(5).

In making this unification it is
necessary to allow transformations between (what were previously thought
to be) particles and antiparticles of the same chirality,
and also between quarks and leptons.  It is
convenient to work with left-handed fields only; since the
conjugate of a right-handed field is left-handed, we don't lose track of
anything by doing so, once we disabuse ourselves of the idea that a
given field is intrinsically either genuine or ``anti''.

As shown in Figure 2, there is one group of ten
left-handed fermions that have
all possible combinations of one unit of each of two different colors, and
another group of five left-handed fermions that each carry
just one negative unit of some
color.  (These are the ten-dimensional antisymmetric tensor and the
complex conjugate of the five-dimensional vector
representation, commonly  referred to as the ``five-bar''.)  What is
important for you
to take away from this discussion
is not so much the precise details of the scheme,
but the idea that {\it the structure of the standard model, with the
particle assignments gleaned from decades of experimental effort and
theoretical interpretation, is perfectly reproduced by a
simple abstract set
of rules for manipulating symmetrical symbols}.  Thus for example
the object
RB in this Figure has just the strong, electromagnetic, and
weak interactions we expect of the complex
conjugate of the right-handed up-quark, without our having to instruct
the theory further.  If you've never done it I heartily recommend
to you the
simple exercise of working out the hypercharges of the objects in
Figure 2 and checking against what you
need in the standard model
-- after doing it, you'll find it's impossible
ever to look at the standard model
in quite the same way again.
\bigskip

\parskip=0pt{
\undertext{SU(5):  5 colors RWBGP}

$\underline{10}$: 2 different color labels (antisymmetric tensor)

$$\matrix{\rm u_L:&\rm RP,&\rm WP,&\rm BP\cr
\rm d_L:&\rm RG,&\rm WG,&\rm BG\cr
\rm u{^c_L}:&\rm RW,&\rm WB,&\rm BR\cr
&\rm (\bar B)&\rm (\bar R)&\rm (\bar W)\cr
\rm e{^c_L}:&\rm GP&&\cr
&(\ )&&\cr
}
\pmatrix{0&\rm u^c&\rm u^c&\rm u&\rm d\cr
&0&\rm u^c&\rm u&\rm d\cr
&&0&\rm u&\rm d\cr
&*&&0&\rm e\cr
&&&&0\cr}$$

$\underline{\bar 5}$: 1 anticolor label

$$\matrix{\rm d{^c_L}:&\rm \bar R,&\rm  \bar W,&\rm  \bar B\cr
\rm e_L:&\rm \bar P&&\cr
\nu_{\rm L}:&\rm \bar G&&\cr
}
\matrix{\rm (d^c&\rm d^c&\rm d^c&{\rm e}&\nu)\cr}$$
\def\boxtext#1{%
\vbox{%
\hrule
\hbox{\strut \vrule{} #1 \vrule}%
\hrule
}%
}
\centerline{
\vbox{\offinterlineskip
\hbox{\boxtext{\rm Y $= -{1\over 3}$ (R+W+B) $+{1\over 2}$ (G+P)}}
}}
\bigskip
\centerline{\bf FIGURE 2}
\medskip
\noindent{\it Figure 2 - Unification of fermions in SU(5).}
\bigskip}

Although it would be inappropriate to elaborate the necessary group theory
here, I'll mention that there is a beautiful extension of SU(5) to
the slightly larger group SO(10), which permits one to unite
all the fermions
of a family into a single multiplet [\geo ].  In fact the relevant
representation for the fermions is a 16-dimensional spinor representation.
Some of its features are depicted in Figure 3.

\parskip=0pt{
$$(\pm \pm \pm \pm \pm)\ \ :\ \ \underline{\rm even}\ \  \# \  of\  -$$
$$10:\matrix{(++-|+-)&6&\rm (u_L,d_L)\cr
(+--|++)&3&\rm u{^c_L}\cr
(+++|--)&1&\rm e{^c_L}\cr}$$

$$\bar 5:\matrix{(+--|--)&\bar 3&\rm d{^c_L}\cr
(---|+-)&\bar 2&{\rm (e_L},\nu_L)\cr}$$

$$1:\matrix{(+++|++)&1&\rm N_R\cr}$$

\centerline{\bf FIGURE 3}
\medskip
\noindent{\it Figure 3 - Unification of fermions in SO(10).
The rule is that all possible combinations of 5 + and - signs occur,
subject
to the constraint that the total number of - signs is even.  The
SU(5) gauge bosons within SO(10) do not change the numbers of signs,
and one see the SU(5) multiplets emerging.  However there are additional
transformations in SO(10) but not in SU(5), which allow any fermion to
be transformed into any other.}
\bigskip}

In addition to the
conventional quarks and leptons the SO(10) spinor contains an
SU(3)$\times$SU(2)$\times$U(1) singlet.  The corresponding
particle has neither strong, weak nor electromagnetic interactions.
It plays an important role in the theory of neutrino masses -- but
that is the topic for another speaker.
Larger gauge groups are also possible.  The exceptional group E(6)
appears naturally in some large classes of
superstring models [\gre ].  The fermions are then
found in multiplets containing a lot of excess baggage that
must be explained away.

\bigskip
{\bf Unification: coupling values}
\medskip

We have seen that simple unification schemes are successful at the
level of
{\it classification}; but new questions arise when we consider the
dynamics which underlies them.

Part of the power of gauge symmetry is that it fully dictates the
interactions of the gauge bosons, once an overall coupling constant
is specified.  Thus if SU(5) or some higher symmetry were exact, then
the fundamental
strengths of the different color-changing interactions would have
to be equal, as would the
(properly normalized) hypercharge coupling strength.  In reality the
coupling strengths of the gauge bosons in SU(3)$\times$SU(2)$\times$U(1)
are not observed to be equal, but rather follow the pattern
$g_3 \gg g_2 > g_1$.

Fortunately, experience with QCD emphasizes that couplings
``run''.  The physical mechanism of this effect is that in quantum field
theory the vacuum must be regarded as a polarizable medium, since
virtual particle-anti-particle pairs can screen charge.  Thus one might
expect that effective charges measured at shorter distances, or equivalently
at larger energy-momentum or mass scales, could be different from what they
appear at longer distances.  If one had only screening then the effective
couplings would grow at shorter distances, as one penetrated deeper insider
the screening cloud.  However it is a famous fact
[\afgt ] that due to paramagnetic
spin-spin attraction of like charge vector gluons [\para ],
these particles tend
to {\it antiscreen} color charge, thus giving rise to
the opposite effect -- asymptotic freedom --
that the effective coupling tends to shrink at short distances.  This
effect is the basis of all perturbative QCD phenomenology, which is a vast
and vastly successful enterprise.  For our present purpose of understanding
the disparity of the observed couplings, it is just what the doctor
ordered.   As was first pointed out by Georgi, Quinn, and Weinberg [\gqw ],
if a
gauge symmetry such as SU(5) is spontaneously broken at some very short
distance then we should not expect that the effective couplings probed at
much larger distances, such as are actually measured at practical
accelerators, will be equal.  Rather they will all have have been affected
to a greater or lesser extent by vacuum screening and anti-screening,
starting from a common value at the unification scale but then diverging
from one another.  The pattern $g_3 \gg g_2 > g_1$ is just what one should
expect, since the antiscreening or asymptotic freedom effect is more
pronounced for larger gauge groups, which have more types of virtual
gluons.

\bigskip\bigskip\bigskip\bigskip\bigskip\bigskip\bigskip\bigskip\bigskip
\bigskip\bigskip\bigskip\bigskip\bigskip\bigskip\bigskip\bigskip\bigskip
\centerline{\bf FIGURE 4}
\noindent{\it Figure 4 - The failure of the running couplings, normalized
according
to SU(5) and extrapolated taking into account
only the virtual exchange of the
``known'' particles of the standard model (including the
top quark and Higgs boson) to meet.  Note that only with quite recent
experiments [\ccexpt ], which greatly improved the precision of the
determination of
low-energy couplings, did the discrepancy become significant.}
\bigskip
The marvelous thing is that the running of the couplings
gives us a truly quantitative
handle on the ideas of unification, for the following reason.  To fix
the relevant aspects of unification, one basically needs
only to fix two parameters: the scale at which the couplings unite, which
is essentially the scale at which the unified symmetry breaks; and
their value when then unite.  Given these, one calculates three outputs:
the three {\it a priori\/} independent couplings for the gauge groups
in SU(3)$\times$SU(2)$\times$U(1).   Thus the framework is eminently
falsifiable.  The miraculous thing is, how close it comes to working
(Figure 4).

The unification of couplings occurs at a very large mass scale,
$M_{\rm un.} \sim 10^{15}~{\rm Gev}$.  In the simplest version, this
is the magnitude of the scalar field vacuum expectation value that
spontaneously breaks SU(5) down to the
standard model symmetry SU(3)$\times$SU(2)$\times$U(1),
and is analogous to the scale $v \approx 250~ {\rm
Gev}$ for electroweak symmetry breaking.  The largeness of
this large scale mass scale
is important in several ways.

$\bullet~$ It
explains why the exchange of gauge bosons that are in SU(5) but not
in SU(3)$\times$SU(2)$\times$U(1), which reshuffles
strong into weak colors
and generically violates the conservation of baryon number,
does not lead to a catastrophically quick decay of nucleons.  The rate
of decay goes as the inverse fourth power of the mass of the exchanged
gauge particle, so the baryon-number violating processes are predicted to
be far slower than ordinary weak processes, as they had better be.

$\bullet~$ $M_{\rm un.}$ is significantly smaller than the Planck scale
$M_{\rm Planck} \sim 10^{19}~{\rm Gev}$ at which exchange of gravitons
competes quantitatively with the other interactions, but not ridiculously
so.  This indicates that while the unification of couplings calculation
itself is probably safe from gravitational corrections, the unavoidable
logical next step in unification must be to bring gravity into the mix.

$\bullet~$ Finally one must ask how the tiny ratio of
symmetry-breaking mass scales $v/M_{\rm un.} \sim 10^{-13}$
required arises dynamically, and whether it is stable.  This is the
so-called gauge hierarchy problem, which we shall discuss in a
more concrete
form a little later.

The success of the GQW calculation in
explaining the observed hierarchy $g_3 \gg g_2 > g_1$ of
couplings and the approximate stability of the proton is quite
striking.
In performing it, we assumed that the known and
confidently expected particles of the standard model exhaust
the spectrum up to the unification scale, and that the
rules of quantum field
theory could be extrapolated without alteration
up to this mass scale -- thirteen orders
of magnitude beyond the domain they were designed to describe.
It is a triumph for minimalism, both existential and conceptual.

(By the way I would like to remark that the running of couplings
calculation, although not at all difficult to do in quantum
field theory, is technically difficult to formulate directly in string
theory.   The loop expansion in gauge couplings comes about through
addition of the contributions from worldsheets of different topology, so
that the ordinary renormalization process involves a coarse-graining over
topologically distinct structures.  Also in field theory the calculation
is most
conveniently formulated ``off shell'', that is using the concept of
virtual particles, which is awkward at best in existing formulations of
string theory.   It is quite disturbing that such a central, physically
transparent calculation should be so troublesome: clearly, some new tools
need to be designed.)

However, on further examination
it is not quite good enough.  Accurate modern measurements
of the couplings show a small but definite discrepancy between the
couplings, as appears in Figure 4.  And heroic dedicated experiments to
search for proton decay did not find it [\pdec ]; they currently
exclude the minimal SU(5) prediction
$\tau_p \sim 10^{31}~{\rm yrs.}$ by about two orders of magnitude.

Given the magnitude of the extrapolation
involved, perhaps we should not have
hoped for more.
There are several perfectly plausible bits of physics
that could upset the calculation, such as the existence of particles
with masses much higher than the electroweak but much smaller than the
unification scale.  As virtual particles these would affect the running
of the couplings, and yet one
certainly cannot exclude their existence on direct experimental
grounds.  If we just add particles in some haphazard
way things will
only get
worse: minimal SU(5) nearly works, so the generic perturbation
from it will be deleterious.  This is a major difficulty for so-called
technicolor models, which postulate many new particles in complex patterns.
Even if some {\it ad hoc\/}
prescription could be made to work,
that would be a disappointing outcome from what
appeared to be one of our most precious, elegantly
straightforward clues regarding physics well
beyond the standard model.

\bigskip
{\bf Virtual supersymmetry?}
\medskip

Fortunately, there is a theoretical idea which is attractive in many
other ways, and seems to point a way out from this impasse.  That is
the idea of supersymmetry [\ferr ].  Supersymmetry is a symmetry that extends
the Poincare symmetry of special relativity
(there is also a general relativistic version).  In a supersymmetric
theory one has not only
transformations among particle states with different energy-momentum but
also between particle states of different {\it spin}.  Thus spin 0
particles can be put in multiplets together with spin ${1\over 2}$
particles, or spin ${1\over 2}$ with spin 1, and so forth.

Supersymmetry is certainly not a symmetry in nature: for example, there
is certainly no bosonic particle with the mass and charge of the electron.
More generally if one defines the $R$-parity quantum number
$$
R~\equiv~ (-)^{3B+L+2S}~,
$$
which should be accurate to the extent that baryon and lepton number are
conserved, then one finds that all currently known particles are
$R$ even whereas their supersymmetric partners would be $R$ odd.
Nevertheless there are
many reasons to be interested in supersymmetry, of which I shall mention
three.

$\bullet$ You will notice that we have made progress in uniting the
gauge bosons with each other, and the various quarks and leptons with each
other, but not the gauge bosons with the quarks and leptons.  It takes
supersymmetry -- perhaps spontaneously broken -- to make this feasible.

$\bullet$ Supersymmetry was invented in the context of string theory, and
seems to be necessary for constructing consistent
string theories containing gravity
(critical string theories) that are at all realistic.
\endpage
{}.
\bigskip\bigskip\bigskip\bigskip\bigskip\bigskip
\centerline{\bf FIGURE 5}
\noindent{\it Figure 5 - A typical quadratically divergent contribution to the
(mass)$^2$ of the Higgs boson, and the supersymmetric contribution which,
as long as supersymmetry is not too badly broken, will largely cancel it.}
\bigskip
$\bullet$ Most important for our purposes, supersymmetry can help
us to understand
the vast disparity between weak and unified symmetry breaking
scales mentioned above.  This disparity is known as the
gauge hierarchy problem.  It actually raises several distinct
problems,
including the following.
In calculating radiative corrections to the
(mass)$^2$ of the Higgs particle from diagrams of the type shown
in Figure 5
one finds an
infinite, and also large, contribution.  By this I mean that
the divergence is quadratic in the ultraviolet cutoff.  No ordinary
symmetry will make its coefficient vanish.
If we imagine that the unification scale provides the cutoff, we find
that the radiative correction to the (mass)$^2$ is much larger than the
final value we want.
(If the Higgs field were
composite, with a soft form factor, this problem might be ameliorated.
Following that road leads to technicolor,
which as mentioned before seems to lead
us far away from our best source of inspiration.)
As a formal matter one can simply cancel the radiative
correction against a large
bare contribution of the opposite sign, but in the absence of some
deeper motivating
principle this seems to be a horribly ugly procedure.
Now in a supersymmetric theory for any set of virtual particles circulating
in the loop there will also be another graph
with their supersymmetric partners circulating.  If the partners were
accurately degenerate, the contributions would cancel.  Otherwise, the
threatened quadratic divergence will be cut off only at virtual momenta
such
that the difference in (mass)$^2$ between the virtual
particle and its supersymmetric partner is negligible.  Thus we will
be assured adequate cancellation if and
only if supersymmetric partners are not
too far split in mass -- in the present context, if the splitting is not
much greater than the weak scale.  This is (a crude version of) the most
important {\it quantitative\/} argument which suggests the relevance
of ``low-energy'' supersymmetry.

The effect of low-energy supersymmetry on the running of the couplings
was first considered long ago [\srun ],
well before the crisis described at the
end of the
previous section was evident.
One might fear that such a huge expansion of the theory, which essentially
doubles the spectrum, would utterly destroy the approximate success of
the minimal SU(5) calculation.  This is not true, however.  To a first
approximation since supersymmetry is a space-time rather than an internal
symmetry it does not affect the group-theoretic structure of the
calculation.  Thus to a first approximation the absolute
rate at which the couplings run
with momentum is affected, but not the relative rates.  The main effect
is that the supersymmetric partners of the color gluons, the gluinos,
weaken the asymptotic freedom of the strong interaction.  Thus they
tend to
make its effective
coupling decrease and approach the others more slowly.  Thus
their merger requires a longer lever arm,
and the scale at which the couplings meet increases by an order of
magnitude or so, to about 10$^{16}$ Gev.
Also the common value of the effective couplings at
unification is slightly larger than in conventional unification
(${g^2_{\rm un.} \over 4\pi } \approx {1\over 25}$ {\it versus\/}
${1\over 40}$).  This increase in unification scale
significantly reduces the predicted rate for proton decay
through exchange of the dangerous color-changing gauge bosons,
so that it no longer conflicts with existing
experimental limits.

Upon more careful examination there is another effect of low-energy
supersymmetry on the running of the couplings, which although quantitatively
small has become of prime interest.  There is an important exception to
the general rule that adding supersymmetric partners does not immediately
(at the one loop level)
affect the relative rates at which the couplings run.  This
rule works for particles that come in complete SU(5) multiplets, such as
the quarks and leptons (which, since they don't
upset the full SU(5) symmetry, have basically no effect)
or for the supersymmetric partners of the
gauge bosons, because they just renormalize the existing,
dominant effect of the
gauge bosons themselves.  However there is one peculiar additional
contribution, from the supersymmetric partner of the Higgs doublet.
It affects only the weak SU(2) and hypercharge U(1) couplings.
(On phenomenological grounds the
SU(5) color triplet partner of the Higgs doublet must be extremely massive,
so its virtual exchange is not important below the unification scale.
{\it Why\/}
that should be so, is another aspect of the hierarchy problem.)
Moreover, for slightly technical reasons even in the
minimal supersymmetric model it is necessary to have
two different Higgs doublets with opposite hypercharges.  The net affect of
doubling the number of Higgs fields and including their supersymmetric
partners is a sixfold enhancement of the asymmetric
Higgs field contribution to
the running of weak and hypercharge couplings.  This causes a
small, accurately calculable change in the calculation.
{}From Figure 6 you see that it is a most welcome one.   Indeed,
in the minimal
implementation of supersymmetric unification, it puts the running of
couplings calculation right back on the money [\scalc ].

Since the running of the couplings with scale is logarithmic the
unification of couplings calculation is not
terribly
sensitive to the exact scale at which supersymmetry is broken,
say between 100 Gev and 10 Tev.  There have
been attempts to push the calculation further, in order
to address this question of
the supersymmetry breaking scale, but they are controversial.  It is not
obvious to me
that such calculations will ever achieve the resolution of interest.
For example, comparable uncertainties arise from the
splittings among the
very large number of particles with masses of order the unification scale,
whose theory is poorly developed and unreliable.
\endpage
{}.
\bigskip\bigskip\bigskip\bigskip\bigskip\bigskip
\bigskip\bigskip\bigskip\bigskip\bigskip\bigskip\bigskip\bigskip\bigskip
\centerline{\bf FIGURE 6}

\noindent{\it Figure 6 - When the exchange of the virtual particles necessary
to implement low-energy supersymmetry, a calculation along the lines
of Figure 4 comes into adequate agreement with experiment.}
\bigskip

In any case, if we are not too greedy
the main points still shine through:

$\bullet$ If supersymmetry is to fulfill its destiny
of elucidating the hierarchy problem in any straightforward way,
then the supersymmetric partners of
the known particles cannot be much heavier than the SU(2)$\times$U(1)
electroweak breaking scale, \ie\ they
should not be beyond the expected reach of SSC.

$\bullet$ If we
assume this to be the case
then the meeting of the couplings takes place in
the simplest minimal models of unification, without further assumption
-- a most remarkable and non-trivial fact.

Thus there are, in my opinion, very good
specific reasons to be hopeful about
the future of experimental
particle physics, if we can summon up the national or
international will to pursue it.

One can build on these ideas in several directions.
The lightest R-odd particle should be stable on cosmological scales
and provides an excellent candidate for the missing matter of
cosmology.  Supersymmetry provides new mechanisms for
CP violation, proton decay, and flavor-changing processes that could come
in at experimentally detectable levels.

There are also important questions ``beyond the standard model'' which the
line of thought I've presented here does not touch.  Among the most
obvious
are the question of why there are three families and why the masses of
fermions and
mixings among them are what they are, and the question of
whether and how gravity might
be united with the other interactions.
Superstring theory undoubtedly provides the most promising and
substantial tools for an assault on these questions.
Unfortunately
nothing nearly as concrete and successful as the line of argument
developed above has emerged from the very extensive and complicated work
on this subject done so far.  On the other hand it is important
that the simplest, apparently
successful
``semi-phenomenological'' ideas about unification, as discussed above,
are not inconsistent with what is known about superstring theory
-- a highly non-trivial fact, since that theory is tightly constrained.

I hope I've been able to convey to you a few core ideas for
physics beyond the standard model that can be understood fairly simply
and that appear likely to be of permanent
value.

\endpage

\refout

\endpage

\end


{\it 1. Unification of interactions and of couplings}

yyRG in string theory

{\it 2. Low energy supersymmetry}

{\it 3. Proton decay}

{\it 4. Hidden matter 1: photinos}

{\it 5. Hidden matter 2: axions}

{\it 6. Neutrino masses}

yyrole in baryogenesis?

{\it 7. Regularities among fermions masses}

yydiscrete symmetries

yyrelation to forbidden processes

{\it 8. Fixed point top mass}

{\it 9. Extra CP violation}

yydipole moments

{\it 10. Ultralight particles}

yydecays and forces

yySTEP project

\end